\documentclass[aps,eqsecnum,prl,twocolumn]{revtex4}
\usepackage{graphics,graphicx}
\usepackage{amsmath}
\usepackage{amssymb,latexsym,mathrsfs}
\usepackage{hyperref}

\def\bea{\begin{eqnarray}}
\def\eea{\end{eqnarray}}
\def\ba{\begin{array}}
\def\ea{\end{array}}

\def\beq{\begin{equation}}
\def\eeq{\end{equation}}

\begin{document}

\title{Quatum Thermodynamics and Coherence in Ion channels}

\author{Samyadeb Bhattacharya$^{1}$ \footnote{sbh.phys@gmail.com}, Sisir Roy$^{2} $ \footnote{sisir@isical.ac.in}}
\affiliation{$^{1,2}$Physics and Applied Mathematics Unit, Indian Statistical Institute, 203 B.T. Road, Kolkata 700 108, India \\}

\vspace{2cm}
\begin{abstract}

\vspace{1cm}

We showed that quantum mechanical superposition can sustain in the process of ion transfer in protein membrane for a substantial period in spite of the presence of the interactions with environmental modes of molecular vibration. The spectral temperature, as defined in quantum thermodynamical framework plays a significant role in maintaining the coherence. The ratio of decoherence time and dwell time has been calculated, which can be directly related to the degree of coherence. The results shead new light to build quantum information system of entangled ionic states in the voltage gated biological channels.

\vspace{2cm}


\vspace{1cm}
\end{abstract}

\vspace{1cm}

\maketitle

The approaches towards dynamics of ion transport in protein membranes (ion channels) are generally considered to be classical and based on molecular dynamics or Brownian dynamics. Recently, one of the present authors showed that \cite{1a} the dynamics of K-ion channel can be explained using nonlinear Schrodinger equation which is compatible with the results of MacKinnon's experimental observation \cite{13}. Here, the two K-ions may form an entangled state within selectivity filter during a finite period of time. The temperature within the channel is generally considered to be high enough to destroy the coherence within very short period. Moreover, molecular modes of the protein environment induce dynamical decoherence, which destroys the quantum mechanical superposition of states in a very short period of time. So quantum mechanical approach in these area of research was mostly speculative and far from experimental realization. But in recent times, some experimental demonstration of the presence of quantum coherence in the process of photosynthetic energy transfer \cite{1,2} lead us to reconsider the theoretical approach and understanding. Here we are concerned with the question that whether and under what condition a sustainable quantum superposition is achievable in the process of transfer of ions through biological channels. Now it is quite a practical argument that the quantum state of the traversing ion is strongly coupled with the molecular vibrational modes of the protein environment and hence fast decoherence is almost absolutely unavoidable. But it is to be noted that the traversal time of the ion through the membrane is also quite small and it is the ratio of the time scale of decoherence with this traversal time that plays an important role in understanding the maintenance of coherence. If the decoherence time is larger than the traversal time of the ion, then the quantum superposition of the ionic states is sustainable enough for the traversing entity within the period of ionic transfer. Here we also need to consider the effect of temperature in estimating the decoherence time. With the recent developments of quantum thermodynamics \cite{11}, a new concept of temperature (known as spectral temperature) for micro-states at non-equilibrium condition has been proposed, instead of the usual concept of thermodynamic temperature. Since ion transport through protein membrane is essentially a non-equilibrium phenomena, we propose that the spectral temperature plays important role in understanding the coherence in the channel dynamics.

Considering the master equation of the density operator for a certain quantum system, the decoherence time \cite{2a} can be written as
\beq\label{1}
\tau_{dec}=\frac{\hbar^2}{2m\gamma K_B T (x-x')^2}
\eeq
where $\gamma$ is the relaxation (dissipation) parameter, $T$ is the thermodynamic temperature and $\Delta x= x-x'$ is the spatial shift of the particle. Now there are numerous definition of traversal time, among which phase time and dwell time are generally accepted by the community \cite{3}. Phase time is equated to dwell time with an additive self-interference term and argued to be the time interval between the energy storage and release in the barrier region \cite{4,5,7,8}. So for non-dispersive barrier, where the self-interference term vanishes, the dwell time can be readily interpreted as the life time of energy storage and release in the barrier region. In a previous work \cite{9}, we have calculated the weak value of dwell time as
\beq\label{2}
\tau_D= \frac{1}{\gamma} \coth\left(\frac{\gamma\tau_M}{2}\right)
\eeq
where $\tau_M$ is the measurement time. \\
The protein membrane can be assumed as an array of molecules with certain modes of vibration. We have mentioned earlier that the quantum states of the traversing entity is interactively coupled to the molecular vibration modes of the protein membrane. The traversing quantum entity loses energy to the vibrational modes, due to the presence of this environmental coupling. So the protein membrane can be interpreted as sort of capacitive system, which can store and release energy with the dwell time as it's lifetime of energy storage. In this case the time interval between the opening and closing of the channel gates can be taken as the measurement time ($\tau_M$). Now we consider a double well potential of the form \cite{10}
\beq\label{3}
V(x)=\frac{1}{2}m\omega^2 x^2 \left[\left(\frac{x}{a}\right)^2-A\left(\frac{x}{a}\right) +B\right]
\eeq
where $A$ and $B$ are dimensionless constants.For the particular case of asymmetric double well, $A=14$ and $B=45$. The bistable nature of the potential is useful in practical situation, since the ionic transfer in the channel can be interpreted as tunneling between the two stability regions situated at $x_0$ and $x_2$ (see fig.1). \\
\begin{figure}[htb]
{\centerline{\includegraphics[width=7cm, height=5cm] {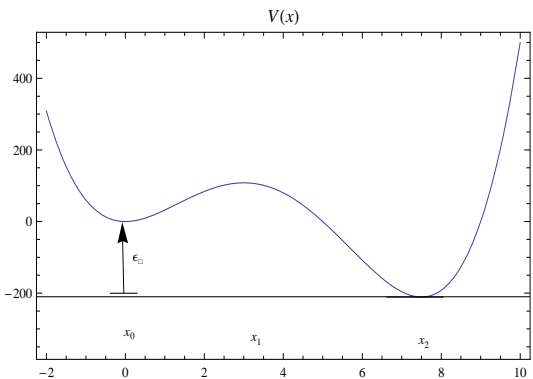}}}
\caption{V(x) vs. x with parameters $A=14$ and $B=45$}
\label{figVr}
\end{figure}
It is also very important to reconsider the concept of temperature in this aspect. Temperature, as we know from the context of thermodynamics, is a property of equilibrium. The ensemble mean of the kinetic energy equals to Boltzman constant times the temperature. So assuming ergodicity, temperature is defined as the time averaged kinetic energy. But the process of ion transfer through channels is a dynamical process interactive to the protein environment and subject to energy exchange with the environment. This is certainly not a situation for thermodynamic equilibrium and hence the usual concept of thermodynamic temperature may not be suitable. Here we introduce the concept of spectral temperature, originally formulated by Gemmer et.al \cite{11}. It is defined as a function of the microstates, to include the non-equilibrium properties of the system. It is formulated as a function of energy occupation probability of the different states of a certain quantum system. This temperature evolves with the evolution of the probability of occurrence. The inverse of the spectral temperature is defined as \cite{11}
\beq\label{4}
\begin{array}{ll}
\frac{1}{K_B T_{spec}}=-\left(1-\frac{P_0 + P_N}{2}\right)^{-1} \sum_{i=1}^{N} \left(\frac{P_{i}+P_{i-1}}{2}\right)\\
~~~~~~~~~~~~~~~\times
\left[\frac{\ln\left(\frac{P_{i}}{P_{i-1}}\right)-\ln\left(\frac{\phi_{i}}{\phi_{i-1}}\right)}{E_i-E_{i-1}}\right]
\end{array}\eeq
where $P_i$ is the probability of finding the particle within an energy compartment with mean energy $E_i$, having the degree of degeneracy $\phi_i$. This definition depends on the energy probability distribution and of course, the spectrum of the concerning system. So it cannot change in time for an isolated system and defined independent of the fact that whether the system is in equilibrium or not. In this definition, the association of quantum probability, which evolves in time, gives temperature a evolving feature representing the dynamical situation between two successive equilibrium. Now here we are approximating the bistable potential as a two-state system having only the ground states of the asymmetric wells. So for our case of non-degenerate two-state system, the expression of the spectral temperature reduces to
\beq\label{5}
\frac{1}{K_B T_{spec}}=-\left(1-\frac{P_0 + P_1}{2}\right)^{-1}  \left(\frac{P_{0}+P_{1}}{2}\right)\left[\frac{\ln\left(\frac{P_{0}}{P_{1}}\right)}{E_0-E_{1}}\right]
\eeq
Where $P_0 ~\mbox{and}~ P_1$ are the probabilities corresponding to the ground states of the lower and higher well respectively. Since $P_0$ is the probability corresponding to the lowest state, it should not decay. Now using the decay probability of the higher state within the time interval equal to the dwell time, we find
\beq\label{6}
\frac{1}{K_B T_{spec}}= \frac{1}{E_1-E_0}\coth\left(\frac{\gamma \tau_M}{2}\right) \coth\left[\frac{1}{2}\coth\left(\frac{\gamma \tau_M}{2}\right)\right]
\eeq
Now if we consider the energy loss by the particle traversing from the higher barrier to the lower one in terms of the usual kinetic temperature ($T_k$) as $E_1-E_0=\frac{1}{2}K_B T_k$, we get
\beq\label{7}
\frac{ T_{spec}}{T_k}= 2\tanh\left(\frac{\gamma \tau_M}{2}\right) \tanh\left[\frac{1}{2}\coth\left(\frac{\gamma \tau_M}{2}\right)\right]
\eeq

\begin{figure}[htb]
{\centerline{\includegraphics[width=7cm, height=5cm] {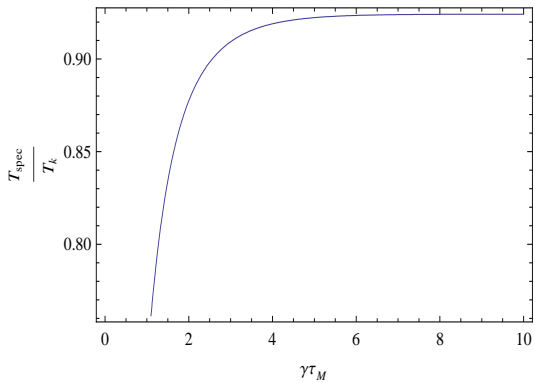}}}
\caption{$T_{spec}/T_{k}$ vs $\gamma\tau_M$. Here we keep $\gamma$ as a constant and basically study the variation with increasing $\tau_M$.}
\label{figVr}
\end{figure}
From eqn (\ref{7}), under the condition $\tau_M\gg \frac{1}{\gamma}$, we find that $T_{spec}\simeq T_k$. ie. the spectral temperature is almost equal to the usual kinetic temperature, if the time interval of gate opening and closing is very greater than the dissipation time scale. If the gate opening and closing mechanism is slow enough to be considered as a quasi-static process, the spectral temperature remains very close to it's kinetic counterpart. Now it should also be noted that though the process is sort of quasi-static one, it should not be considered as a reversible process, because the coupling to the molecular vibrational modes of the protein environment ensures some generation of dissipative entropy.\\
In case of a double well potential given by (\ref{3}), using (\ref{1}),(\ref{2}) and (\ref{7}) we get
\beq\label{8}
\frac{\tau_{dec}}{\tau_D}=\frac{2\hbar}{w}\sqrt{\frac{2}{m \epsilon_0}}\coth\left[\frac{1}{2}\coth\left(\frac{\gamma \tau_M}{2}\right)\right]
\eeq
where $\epsilon_0=E_1-E_0$ is the asymmetry energy of the potential and $w=\frac{15a}{2}$ is the separation length between the wells.
\begin{figure}[htb]
{\centerline{\includegraphics[width=7cm, height=5cm] {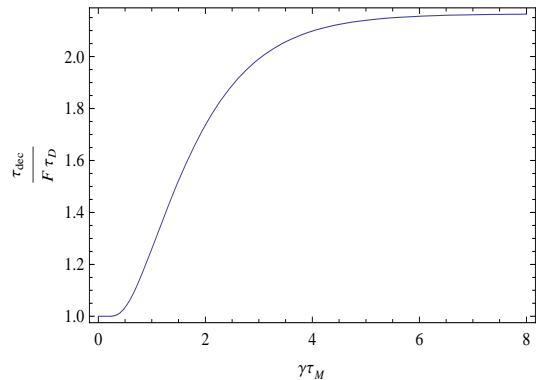}}}
\caption{$\tau_{dec}/\tau_{D}$ vs $\gamma\tau_M$. Here also we keep $\gamma$ as a constant and basically study the variation with increasing $\tau_M$. Here $F=\frac{2\hbar}{w}\sqrt{\frac{2}{m \epsilon_0}}$. As with increment of $\tau_M$, the process becomes quasi-static, we see that the ratio of the two timescales also reach a stable value }
\label{figVr}
\end{figure}
For the quasi-static condition $\tau_M\gg \frac{1}{\gamma}$
\beq\label{9}
\frac{\tau_{dec}}{\tau_D}\simeq \frac{4.5\hbar}{w}\sqrt{\frac{2}{m \epsilon_0}}
\eeq
So we find that in the quasi-static region, whether the decoherence time scale is larger than the dwell time depends on the mass of the traversing particle, the length scale of the channel and the asymmetry energy which is basically the energy lost by the particle during the process of traversal. The ratio of the time scales is inversely proportional to all the mentioned parameters. i.e. greater the inertia of the traversing particle, stronger the interaction (hence greater energy loss) and larger the traversal length, faster shall be the decoherence and hence the situation will be more and more classical. But if the parameter can be chosen in such a way in which the process of decoherence is slower than the process of ionic transfer (i.e. $ \tau_{dec}/\tau_D > 1$), then quantum superposition will be sustainable enough within the time period of ionic transfer. \\
Now we turn our attention to the ``degree of coherence" for such cases of ionic transfer through protein membranes and try to establish a relation of the afore mentioned ratio of decoherence-dwell time scales with it. Convenient way to discuss coherence in quantitative terms can be done through the introduction of normalized form of correlation functions such as ``degree of coherence" \cite{12}
\beq\label{10}
g^{(n)}(\xi_1...\xi_{2n},t_1...t_{2n})= \frac{G^{(n)}(\xi_1...\xi_{2n},t_1...t_{2n})}{\prod_{j=1}^{2n}[G^{(1)}(\xi_j,\xi_j,t_1)]^{1/2}}
\eeq
where $\xi=x/a$. For our case of double well system approximated as two-state system, (\ref{10}) is reduced to
\beq\label{11}
g^{(1)}= \frac{G_{12}^{(1)}(\tau)}{\sqrt{G_{11}^{(1)}(0)G_{22}^{(1)}(0)}}
\eeq
where
\beq\label{12}
\begin{array}{ll}
G_{12}^{(1)}(\tau)=\langle \psi_1(\xi,t)\psi_2(\xi,t+\tau)\rangle\\
~~~~~       = e^{-\gamma \tau}\int_{-\infty}^{\infty} \psi(\xi-\xi_{1})\psi^{*}(\xi)d\xi
\end{array}
\eeq
and
\beq\label{12a}
0\leq g^{(1)}\leq 1
\eeq
For completely coherent situation, the degree of coherence is 1 and for completely decohered case it is 0. Generally the value should lie in between. If we approximate the potential around the left well at $\xi=0$ as a harmonic potential, we find that the wavefunction can be estimated as \cite{10}
\beq\label{13}
\psi(\xi)=\left(\frac{\nu}{\pi}\right)^{1/2}\exp\left[-\frac{1}{2}\nu\xi^2\right]
\eeq
where $\nu=\sqrt{B}\frac{m\omega a^2}{\hbar}$. Expressing the asymmetry energy as
\beq\label{14}
\epsilon_0=V(x_0)-V(x_2)
\eeq
we find that for a double well potential ($A=14$ and $B=45$)
\beq\label{15}
\omega=\frac{4}{15a}\sqrt{\frac{2\epsilon_0}{15m}}=\frac{2}{w}\sqrt{\frac{2\epsilon_0}{15m}}
\eeq
So for the interval of dwell time given by equation (\ref{2}), we find that the degree of coherence
\beq\label{16}
g^{(1)}= \exp\left[-\sqrt{\frac{m\epsilon_0}{2}}\frac{w}{\hbar}\right] \exp\left[-\coth\left(\frac{\gamma \tau_M}{2}\right)\right]
\eeq
For the quasi-static condition $\tau_M \gg 1/\gamma$
\beq\label{17}
g^{(1)}\approx \exp\left[-\sqrt{\frac{m\epsilon_0}{2}}\frac{w}{\hbar}\right]
\eeq
From equation (\ref{9}) and (\ref{17}), we can also establish a relation
\beq\label{18}
\frac{\tau_{dec}}{\tau_D}=\frac{4.5}{\ln\left[\frac{1}{g^{(1)}}\right]}
\eeq
From this relation we see that for completely coherent and decoherent situation, the decoherence-dwell time ratio is infinity and zero respectively and in general situation it lies in between. So this time-scale ratio gives us a certain measure of coherence. As the value of this ratio gets bigger, the ``quantumness" of the traversing entity increases.
\begin{figure}[htb]
{\centerline{\includegraphics[width=7cm, height=5cm] {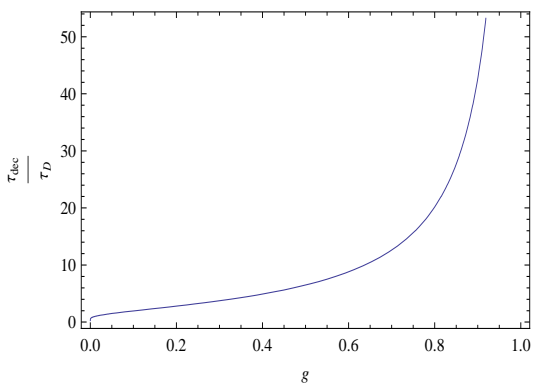}}}
\caption{$\tau_{dec}/\tau_D$ vs. $g^{(1)}$}
\label{figVr}
\end{figure}
\\ Depending on the above analysis, we suggest that the ion selectivity filter may exhibit quantum coherence which can play crucial role in the process of selectivity and conduction of specific ions in biological membranes. For the time scales shorter than that of decoherence time, quantum coherence can be expected to sustain and have vital importance in the dynamics, despite the presence of interactive protein environment. Our analysis shows that for a sort of quasi-static situation, where the gate opening and closing mechanism is slower than the relaxation (dissipation) time scale, the decoherence-dwell time ratio reaches a static value and can also be greater than unity depending on the mass, energy and length parameters. In such situations, coherent phenomena like entanglement can be of vital importance in understanding the mechanism of selectivity and transport. In case of $K+$ filter, there exists two energetically almost degenerate binding states, commonly referred as ($1,3$) and ($2,4$) states \cite{12a,12b,12c}. Presence of quantum superposition may lead us to explain the transport phenomena in terms of quantum mechanical tunneling between these two states. The interplay between quantum coherence and environmental noise induced dephasing may also be of fundamental importance. Progress in the atomic spectroscopy of membrane proteins indicate that protein membrane organization may carry a certain coding potency, implying quantum entanglement within ion channels \cite{12a,12b,12c,12d}. Increasing number of studies are indicating towards the probabilistic nature of ion channel gating mechanism \cite{13,14}. In the light of these researches, we conclude that it may be relevant to build certain model of quantum information system driven by the entangled ionic states in the voltage gated selectivity filters, which can provide necessary inferences into the biological ion channel dynamics.

\end{document}